\documentclass[journal=jpccck]{achemso}


\usepackage[version=3]{mhchem} 
\usepackage{siunitx}
\usepackage{color,soul}
\usepackage{gensymb}
\usepackage{hyperref}
\usepackage{caption}
\usepackage{amssymb}
\author{Zhenzhu Li}
\affiliation{Department of Materials, Imperial College London, London SW7 2AZ, UK}
\alsoaffiliation{Imperial-X, Imperial College London, W12 7SL, UK}%
\alsoaffiliation{Imperial Global Singapore, CREATE Tower, 138602, Singapore}
\email{zhenzhu.li@imperial.ac.uk}

\author{Aron Walsh}%
\affiliation{Department of Materials, Imperial College London, London SW7 2AZ, UK}

\title[]
{Platonic representation of foundation machine learning interatomic potentials}


\begin{document}

%
%

\begin{abstract}
Foundation machine learning interatomic potentials (MLIPs) are trained on overlapping chemical spaces, yet their latent representations remain model-specific. Here, we show that independently developed MLIPs exhibit statistically consistent geometric organisation of atomic environments, which we term the Platonic representation. By projecting embeddings relative to a set of atomic anchors, we unify the latent spaces of seven MLIPs (spanning equivariant, non-equivariant, conservative, and non-conservative architectures) into a common metric space that preserves chemical periodicity and structural invariants. This unified framework enables direct cross-model optimal transport, interpretable embedding arithmetic, and the detection of representational biases. Furthermore, we demonstrate that geometric distortions in this space can indicate physical prediction failures, including symmetry breaking and incorrect phonon dispersions. Our results show that the latent spaces of diverse MLIPs present consistent statistical geometry shaped by shared physical and chemical constraints, suggesting that the Platonic representation offers a practical route toward interoperable, comparable, and interpretable foundation models for materials science.
\end{abstract}


\section{Introduction}
Machine learning interatomic potentials (MLIPs) offer a powerful tool to accelerate the inference of energy (\textit{E}), forces (\textit{F}), and stresses ($\sigma$) in atomistic structures. The development of MLIPs has evolved from early data-fitting approaches using crystal graphs, such as M3GNet\cite{Chen2022}, to recent architectural designs incorporating atom-centred expansions with high body-order, message passing, and equivariance in implementations such as ACE, MACE, NequIP, and SevenNet\cite{Kovács2021, Batatia2025, Batzner2022, Park2024}. This evolution includes the divergence between conservative architectures, where forces are derived as the negative gradient of the energy, and non-conservative regimes that predict forces directly, as explored in Orb-v3\cite{rhodes2025orbv3atomisticsimulationscale, bigi2025darkforcesassessingnonconservative}. While the Matbench Discovery\cite{Riebesell2025} framework provides a benchmark for these state-of-the-art (SOTA) pretrained MLIPs, current evaluations rely on metrics related to the predictive performance of the models. 

Recent work in natural language processing has proposed the ``Platonic representation hypothesis". It posits that neural networks trained with different objectives and modalities converge toward shared statistical models of reality in their representation spaces\cite{pmlr-v235-huh24a}. Whether an analogous phenomenon exists in materials science remains an open question. Unlike language models that learn from human-generated text, MLIPs learn from quantum mechanical calculations that vary in fidelity (e.g. choice of exchange-correlation functional, basis set) and chemical coverage. Nevertheless, the underlying physics and chemistry, governed by quantum mechanics and the fundamental nature of interatomic interactions, provides a ground truth that all models must approximate.

In this work, we demonstrate that foundation MLIPs converge toward a shared, architecture-independent latent geometry, termed the Platonic representation. We introduce an anchor-based projection framework to unify the embeddings of various foundation models, enabling interoperability that is otherwise mathematically inaccessible. Using this unified space, we apply cross-model optimal transport analysis to quantify the geometric distance between pre-trained potentials. Furthermore, we develop an embedding arithmetic scheme that consistently represents materials and reactions across different architectures. Finally, we suggest that deviations from this Platonic geometry can potentially serve as a diagnostic tool, effectively highlighting training divergence and physical inconsistencies.

\section{Results}

\subsection{Aligning incompatible representations}

We chose seven foundation MLIPs representing distinct architectures, datasets, and approaches to equivariance and energy conservation. These include three MACE-MP-0 variants (Large, Medium, and Small)\cite{batatia2025foundationmodelatomisticmaterials} trained on the Materials Project Trajectory Dataset (MPtrj)\cite{Deng2023}; two OMat24-based models\cite{barrosoluque2024openmaterials2024omat24} (MACE-omat and Seven-omat); and two Orb-v3 models (Orb-v3-con-omat and Orb-v3-dir-omat), which differ in their treatment of force conservation. For each model, we extracted 282,847 atomic embeddings across 27,136 structures from the MP-20 dataset\cite{Xie2021CrystalDV}.

We applied principal component analysis (PCA) to project these embeddings into a two-dimensional space, where the first two principal components (PCA1 and PCA2) capture the directions of greatest variance. While we also examined nonlinear visualisation techniques such as uniform manifold approximation and projection (UMAP)\cite{mcinnes2020umapuniformmanifoldapproximation} (Fig. S1), but they distort global geometry and reconfigure relative distances. 
As shown in Fig. 1, the raw embeddings cannot be compared directly. Dimensionality varies by architecture; for instance, MACE embeddings use 128 dimensions, whereas Orb-v3 uses 256. More importantly, the representations depend strongly on the training setup, including the specific architecture, loss function, and random initialisation. Even the MACE-MP-0 variants, which share the same training dataset (MPTrj) and objective, exhibit divergent latent spaces. This indicates that original representations act as arbitrary coordinate systems learned by each model architecture.

\begin{figure}[H]
    \centering
    \includegraphics[width=1\linewidth]{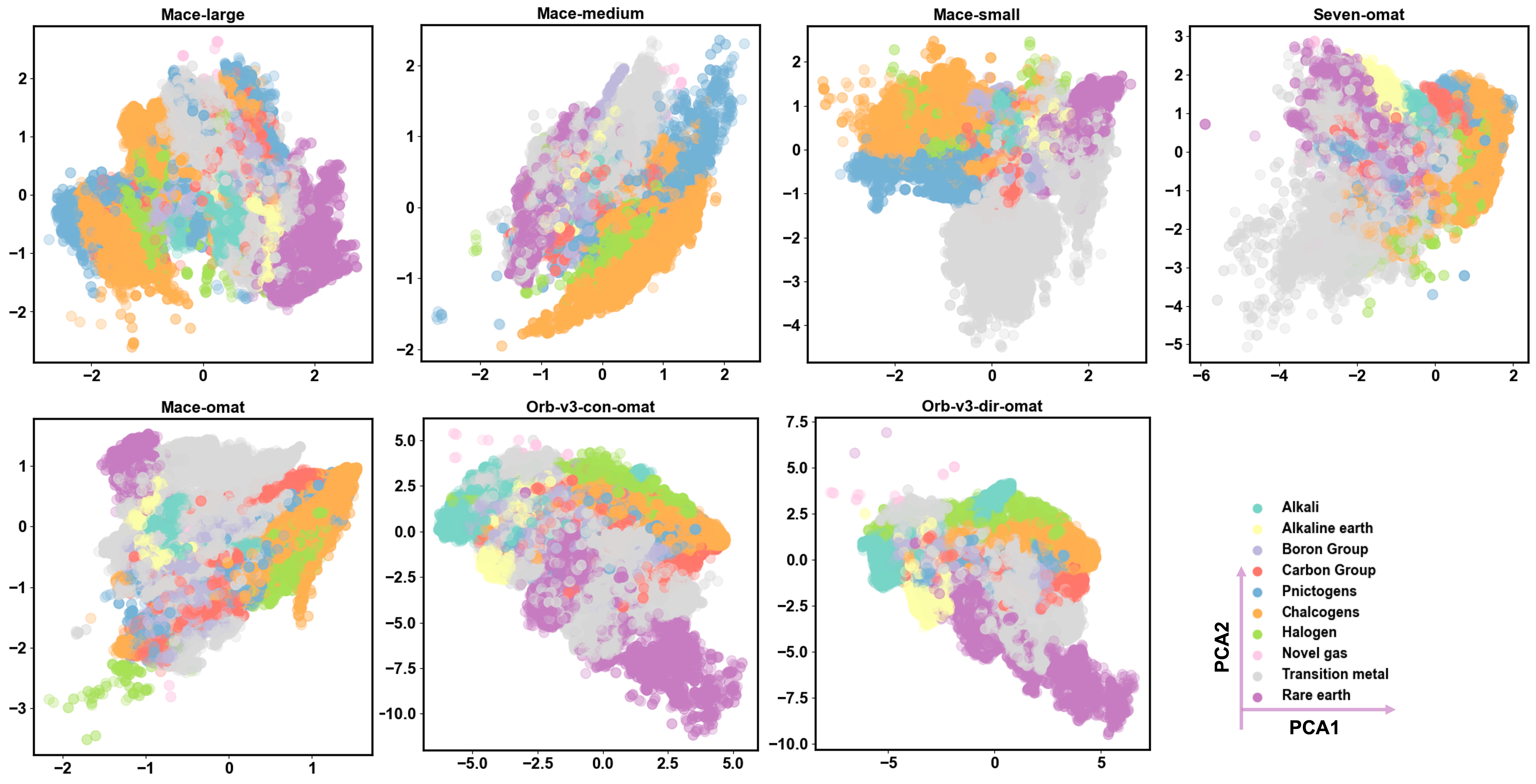}
    \caption{\textbf{Model-specific embeddings are incompatible before transformation.} 2D-PCA projections of atomic embeddings from seven foundation MLIPs reveal distinct variance directions and element clustering patterns. Although all models are trained to predict the same physical quantities for overlapping material sets, they learn embeddings in incompatible coordinate systems.}
    \label{fig:enter-label}
\end{figure}

For alignment, we first project the model-specific embeddings into a unified latent space. We achieve this using the relative representation strategy proposed by Moschella et al.\cite{DBLP:conf/iclr/MoschellaMFNLR23}, establishing a common coordinate frame independent of the original architecture.
We use cosine similarity instead of distance functions (Euclidean $L_2$ or Manhattan $L_1$) as it provides scale invariance—critical when comparing embeddings with varying norms (e.g., 3-5× differences across models; Table 1). This choice naturally bounds projections to [0,1], facilitating interpretation. The transformation framework is illustrated in Fig. 2a. Let $\mathbf{e}_i \in \mathbb{R}^d$ denote the original embedding of an atomic environment $i$. We select a set of $K$ anchor vectors, $\{\mathbf{a}_1, \mathbf{a}_2, \dots, \mathbf{a}_K\}$, from the embedding manifold. The Platonic transformation, $T(\mathbf{e}_i)$, projects $\mathbf{e}_i$ into the anchor-defined space based on its cosine similarity to these reference points:
\begin{equation}
    \mathbf{z}_i = T(\mathbf{e}_i)
    = \left[
        \cos(\mathbf{e}_i, \mathbf{a}_1),
        \cos(\mathbf{e}_i, \mathbf{a}_2),
        \dots,
        \cos(\mathbf{e}_i, \mathbf{a}_K)
      \right]^{\!\top},
\end{equation}
where the cosine similarity is defined as:
\begin{equation}
    \cos(\mathbf{e}_i, \mathbf{a}_k)
    = \frac{\mathbf{e}_i^{\top} \mathbf{a}_k}
           {\|\mathbf{e}_i\|_2 \, \|\mathbf{a}_k\|_2}.
\end{equation}
The resulting vector $\mathbf{z}_i \in \mathbb{R}^K$ constitutes the embedding in the unified Platonic space. By computing the cosine similarity between an input embedding and each anchor, we transform the absolute, model-dependent coordinates into a relative representation. The dimensionality of this new space is determined solely by the number of anchors, $K$, with each axis corresponding to the similarity to a specific anchor, rather than an arbitrary feature channel.

To ensure the anchor set captures the diversity of the embedding manifold, we compared random sampling with DImensionality-Reduced Encoded Clusters with sTratified (DIRECT) sampling\cite{Qi2024}. Originally developed to facilitate MLIP training, DIRECT sampling selects points that maximise coverage in the chemical latent space. We evaluated these strategies across multiple random seeds $\in$\{0, 42, 12345\}. As detailed in Table S1, DIRECT sampling yields significantly more diverse anchor sets, characterised by larger pairwise distances ($>1.5$) and lower Silhouette scores ($<0.1$) compared to random selection. Consequently, we employ the DIRECT strategy throughout this work to ensure broad coverage and high representational diversity.

\begin{figure}[H]
    \centering
    \includegraphics[width=0.75\linewidth]{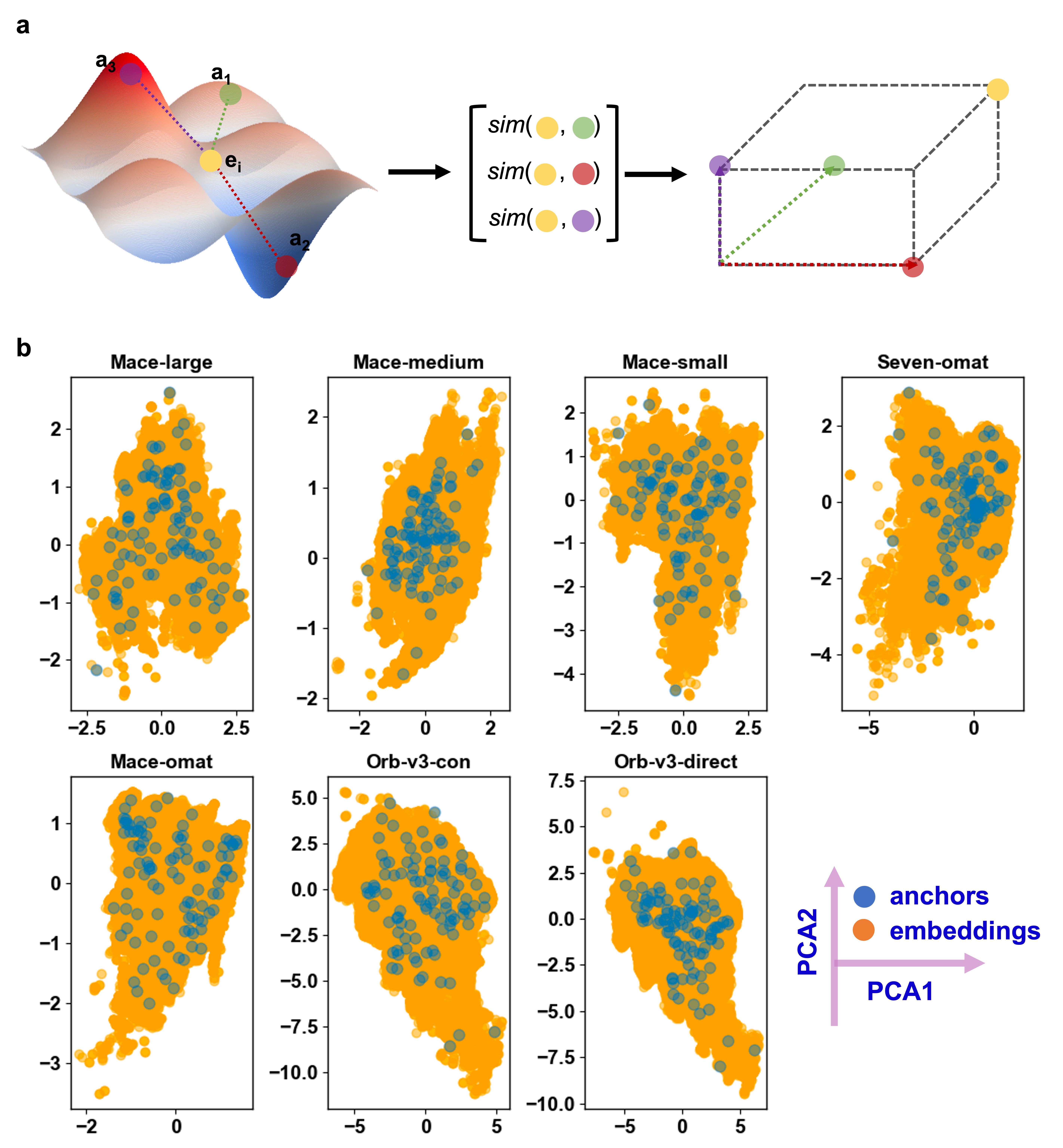}
    \caption{\textbf{Construction of the unified coordinate system.} (a) Schematic of the anchor-based transformation. Green, red, and purple points represent a set of three anchor vectors ($\mathbf{a}_k$), while the yellow point indicates a sample vector $\mathbf{e}_i \in \mathbb{R}^d$. The transformation projects $\mathbf{e}_i$ into the anchor-defined space via cosine similarity. (b) Distribution of 100 DIRECT-sampled anchors (blue dots) overlaid on the PCA projection of the original embedding manifold (orange background).}
    \label{fig:anchor_dist}
\end{figure}

\subsection{Convergence to a shared chemical geometry}

The transformed representations obtained using anchor sets of varying sizes 
($K \in \{3, 8, 20, 50, 100, 200, 400\}$ are presented in Fig.  3a. Even a minimal set of three anchors effectively aligns embeddings across disparate architectures,  highlighting transition metals (grey) forming a distinct cluster, clearly separable from chalcogens (orange), halogens (green), and other main-group elements. As the anchor count increases, the unified representation stabilises, with variance converging once the anchor set size $K$ reaches 100. Here, we display relative representations from Mace-medium, Mace-small, and Orb-v3-con-omat, representing both equivariant and non-equivariant architectures. A more comprehensive trend of convergence across all models is provided in Figs. S2 and S3.

These results indicate that anchors act as stable reference points defining a shared coordinate system. Once the principal geometric relationships are pinned to these anchors, the remaining embeddings naturally align, reflecting the same underlying physical drivers regardless of the model origin. Furthermore, the evolution of the embedding topology with increasing $K$ reveals a hierarchical organisation within the chemical space. While a small number of anchors captures the coarse global structure, increasing the anchor density resolves finer structural details.

A characteristic of the shared representation is the sensitivity to the anchor set. As illustrated in Fig.  3b-c, representations constructed using DIRECT sampling yield a more unified manifold than those using random sampling. DIRECT-sampled anchors produce tighter clustering with mean pairwise distances ranging from 1.58 to 2.67, whereas random sampling leads to skewed, loosely distributed spaces (distances of 1.80–3.11; see Table S2). This sensitivity is particularly pronounced for models trained on the OMat24 dataset (Orb-v3 variants). Their Platonic embeddings exhibit skewed distributions relative to the MACE-MP-0 models trained on MPTrj, particularly when anchors are chosen at random. This behaviour persists across anchor set sizes from $K=20$ to $400$ (Fig.  S2 and S3).

\begin{figure}[H]
\centering
\includegraphics[width=1\linewidth]{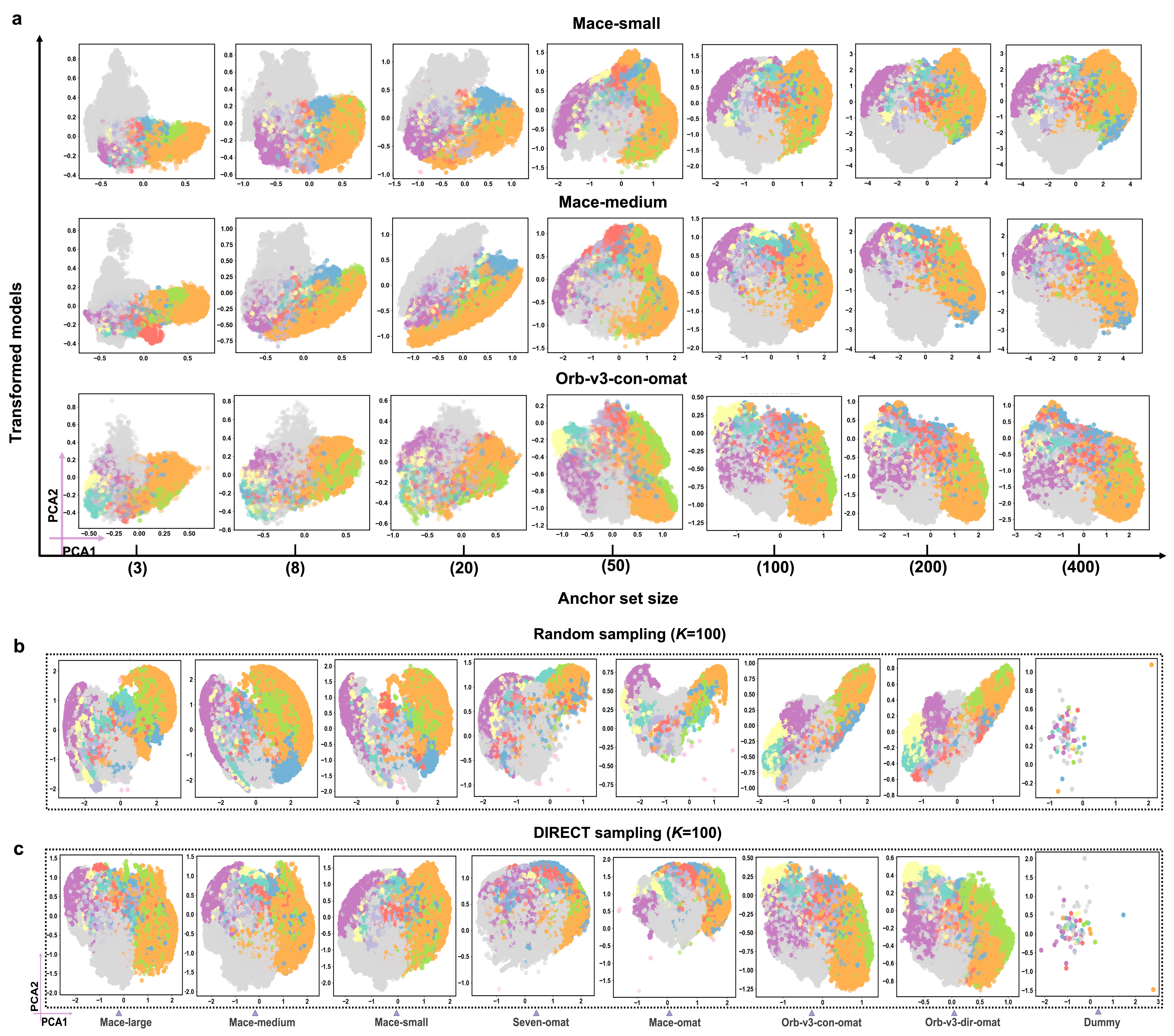}
\caption{\textbf{Variation of transformed representations with anchor set size and sampling strategy.} (a) Transformed representations as a function of anchor set size ($K = 3$ to $400$). (b) 2D PCA projections of converged representations using 100 randomly sampled anchors. (c) Projections using 100 DIRECT-sampled anchors. Despite architectural diversity, all models transformed with DIRECT sampling show substantial alignment. Non-equivariant models (Orb-v3) exhibit systematic skewness. The Dummy model (untrained, random weights) displays no chemical structure, confirming that alignment reflects learned physical knowledge. Colourmap follows the labelling in Figure 1.}
    \label{fig:platonic_rep}
\end{figure}

The role of anchor diversity parallels findings in image representation alignment\cite{pmlr-v97-kornblith19a}. Small, non-diverse anchor sets ($\leq 20$) provide only a coarse global alignment. As the anchor set grows ($100$--$400$), diversity becomes essential to constrain the mapping, correct local distortions, and recover a globally consistent alignment. This is analogous to studies mapping language models to human neural responses\cite{Caucheteux2022}, where the fidelity of the unified space depends on the structural richness of the probes. To validate that this geometry arises from learned physics rather than mathematical artefacts, we applied the transformation to a ``Dummy" MACE model initialised with random weights (see Methods). As shown in the rightmost columns of Fig. 3b-c, this untrained model yields no discernible chemical structure. This contrast confirms that the Platonic representation emerges strictly from meaningful correlations learned from the data. We further demonstrate the universality of this framework by extending it to additional architectures, including NequIP-OAM-L and Mace-mpa-0, in Fig. S5.

\subsection{Quantifying representational interoperability}

To assess the interoperability of embeddings in the unified space, we evaluate global topology, local neighbourhood preservation, and distributional distance. We employ a composite evaluation framework comprising Procrustes analysis ($\mathrm{Score_{Procrustes}}$) to measure global alignment after rotation\cite{Schönemann_1966}, Mutual $k$-nearest neighbours ($\mathrm{Score_{mKNN}}$) to quantify local decision boundary overlaps, and Normalized Optimal Transport ($\mathrm{Score_{OT}}$) to determine the minimal effort required to morph one latent distribution into another\cite{MAL-073}.

As shown in Fig. 4b, foundation MLIPs exhibit substantial global topological similarity, particularly among architectures sharing design principles (e.g., MACE variants show $\mathrm{Score_{Procrustes}} > 0.86$). However, this macro-scale alignment masks significant local divergence. The local neighbourhood similarity ($\mathrm{Score_{mKNN}}$) remains low ($< 0.38$) across all pairs (Fig. 4a), significantly dropping from the values observed in the original, unaligned spaces (Fig. S6). This disparity indicates that while different models converge on the same global physical manifold, their local decision boundaries remain distinct. Notably, non-equivariant models (Orb-v3) show near-zero $\mathrm{Score_{mKNN}}$ against equivariant models, confirming that the lack of symmetry constraints leads to a fundamentally different encoding of local atomic environments.

\begin{figure}[H]
\centering
\includegraphics[width=1\linewidth]{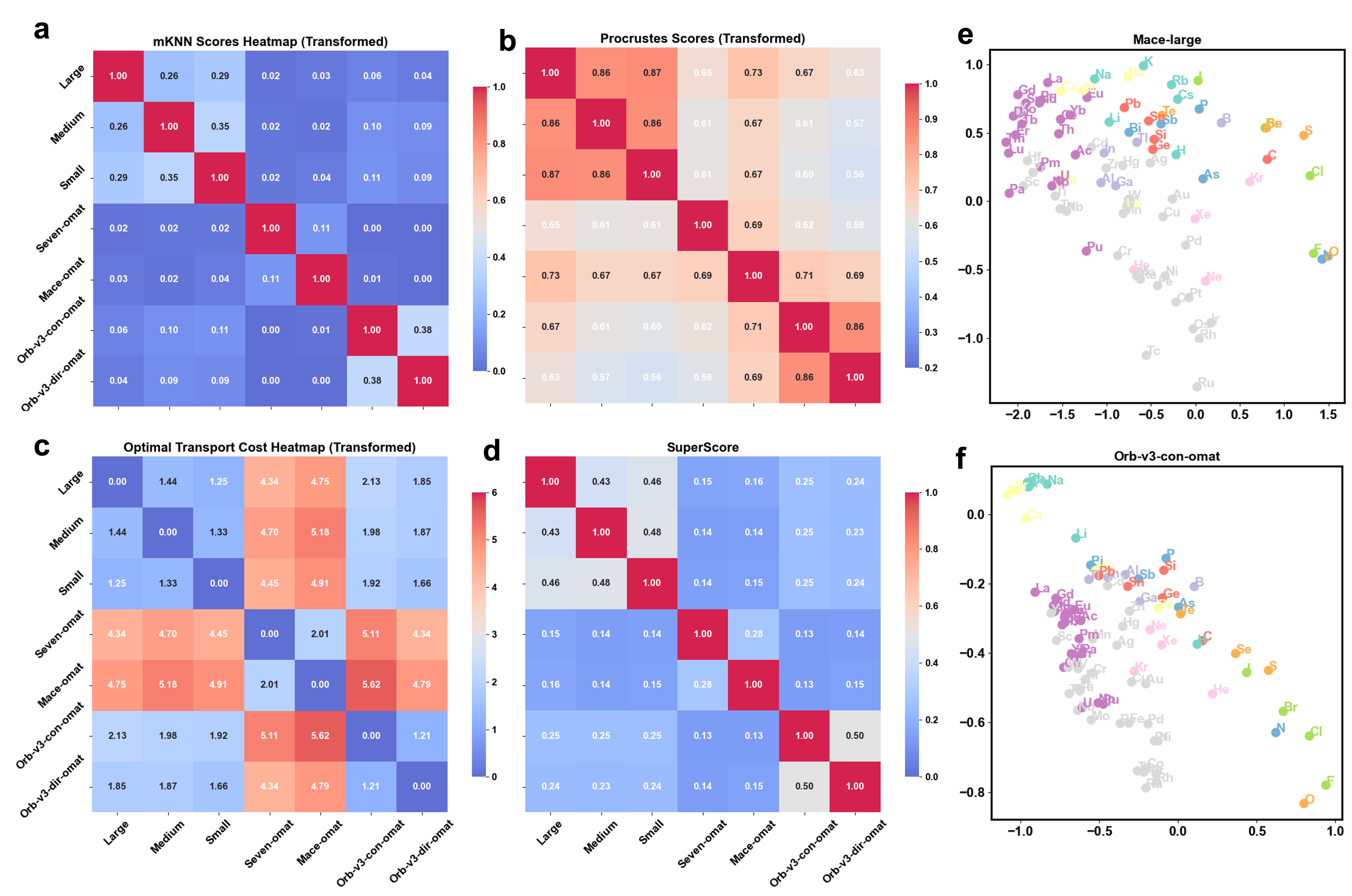}
\caption{\textbf{Quantifying model similarity and chemical preservation.} (a) mKNN scores (local fidelity), (b) Procrustes scores (global alignment), (c) Normalized Optimal Transport (OT) cost, and (d) the composite SuperScore. (e, f) Element-level embeddings projected into the unified space ($K$=100 anchors) reveal consistent periodic clustering across all seven models (Fig. S7), here results with MACE-large (e) and conservative Orb model (f) are presented. Equivariant architectures (MACE, SevenNet) produce compact clusters, whereas non-equivariant models (Orb-v3) exhibit skewness, suggesting equivariance constraints sharpen chemical organization.}
\label{fig:metrics_chemical}
\end{figure}

The OT cost map (Fig. 4c) partitions the seven models into three coherent clusters, revealing the interplay between architecture and data. High OT distances separate the OMat-trained equivariant models (Seven-omat, Mace-omat) from the MACE-MP-0 family, isolating the influence of dataset composition. The largest distances, however, separate the non-equivariant Orb-v3 models from all others, confirming that disparate physical constraints create distinct energy landscapes. We can define a combined metric:
\[
\mathrm{SuperScore} = 
\frac{1}{2}(\mathrm{Score_{Procrustes}} +
     \mathrm{Score_{mKNN}}) \times 
     e^{-{\mathrm{Score_{OT}}}
     }.
\]
that averages the global and local alignment scores, while applying a penalty based on the transport cost. Bounded between [0, 1], SuperScore offers a summary of representational compatibility. 

Beyond statistical metrics, the unified representation explicitly preserves chemical structure (Fig. 4e-f; all models in Fig. S7). When projecting element-level embeddings that are mean-pooled from a unifed representation, all models produce a similar topology of the periodic table. Elements from the same group, such as halogens, chalcogens, pnictogens, cluster coherently, and periodic trends (e.g., F–Cl–Br–I) form smooth trajectories. However, architectural biases persist: equivariant models (MACE, SevenNet) produce compact, spherical clusters, whereas non-equivariant models (Orb-v3) yield skewed distributions. This suggests that without explicit symmetry constraints, the model relies more heavily on statistical correlation than physical symmetry to organise chemical space.

\subsection{Algebraic consistency and model stitching}

A potential utility of a Platonic representation is embedding arithmetic across different models. By mapping into a unified coordinate system, vector operations can be performed. We evaluate this through three case studies: a complex oxide (\ce{Na3MnCoNiO6}), a symmetry-sensitive polymorphic pair of \ce{TiO2}, and a solid-state reaction to synthesise \ce{BaTiO3}. All embeddings are computed in the unified space ($K=100$ anchors), with cross-model similarity (c-sim) measured relative to MACE-large and extended cases are presented in the suplementary information.

We define the material-level embedding, $\mathbf{z}_{\mathrm{Mater}}$, as the centroid of its atomic constituents in the unified space:
\begin{equation}
\mathbf{z}_{\mathrm{Mater}} 
= \frac{1}{N_{\mathrm{atoms}}} 
\sum_{i=1}^{N_{\mathrm{atoms}}} \mathbf{z}_i.
\end{equation} 
As shown in Table 1, models exhibit agreement on $\mathbf{z}_{\mathrm{Mater}}$ for \ce{Na3MnCoNiO6}, with c-sim values consistently between 0.79 and 0.87 (excluding Orb-v3-con-omat). Notably, OMat24-trained equivariant models (Seven-omat, Mace-omat) produce vectors with significantly larger norms ($l \approx 4.5\text{--}5.4$) compared to MACE-MP-0 models ($l \approx 1.1$). This scaling factor ($\sim 3\text{--}5\times$) is consistent across materials (Table S3), indicating that while architectures may vary in signal magnitude, they encode similar angular information relative to the transformed axes.

To assess sensitivity to structural degrees of freedom, we analysed two \ce{TiO2} polymorphs: one orthorhombic (Pbcn) and one tetragonal ($\mathrm{P4_2/mnm}$) space group. While intra-model similarity (i-sim) between polymorphs is high ($\geq 0.95$), the cross-model agreement on the difference vector, $\mathrm{\textbf{z}_{Morph}} = \mathrm{\textbf{z}_{Mater1}} - \mathrm{\textbf{z}_{Mater2}}$, is low ($\leq 0.40$). This suggests that while global material identity is preserved, current foundation models, and by extension their unified representations, smooth over the subtle local distortions that distinguish polymorphs, highlighting a resolution limit in current pooling strategies.

\begin{table*}[h!]
\caption{\textbf{Cross-model embedding arithmetic.} Vector norms ($l$) and cosine similarities (c-sim, relative to MACE-large) for: (i) \ce{Na3MnCoNiO6} material embeddings; (ii) \ce{TiO2} polymorph differences; and (iii) \ce{BaTiO3} formation reaction vectors (standard vs. zero-shot stitched). i-sim denotes intra-model similarity between polymorphs.}
\label{table1}
\setlength{\arrayrulewidth}{0.4mm}
\begin{tabular}{ccccc}
\hline
\textbf{Model} & $\mathbf{ \textbf{z}_{Mater}}$ & $\mathbf{\textbf{z}_{Morph}}$ & $\mathbf{\textbf{z}_{React}}$ & $\mathbf{\textbf{z}_{React-stitch}}$  
\\ 
& ($l$, c-sim) & ($l_1, l_2$, i-sim, c-sim) & ($l$, c-sim) & ($l$, c-sim) \\
\hline
MACE-large   & 1.53, 1.00  & 1.51, 1.59, 0.97, 1.00  & 1.26, 1.00 & 1.39, 1.00 \\
MACE-medium  & 1.05, 0.84  & 1.21, 1.43, 0.95, 0.46  & 1.08, 0.69 & 1.69, 0.92 \\
MACE-small   & 1.14, 0.87  & 1.62, 1.53, 0.97, 0.40 & 0.99, 0.85 & 0.99, 0.88 \\
Seven-omat & 4.50, 0.79 & 5.29, 5.50, 1.00, 0.14 & 5.58, 0.77 & 3.72, 0.38 \\ 
MACE-omat & 5.44, 0.79  & 5.88, 6.15, 1.00, 0.47 & 5.74, 0.75 & 4.33, 0.43 \\ 
Orb-v3-con & 0.68, 0.54 & 1.56, 1.83, 0.99, 0.36 & 1.05, 0.48 & 2.28, 0.82 \\ 
Orb-v3-dir & 1.28, 0.79  & 2.01, 2.30, 0.99, 0.39 & 1.64, 0.73 & 1.90, 0.67 \\
\hline
\end{tabular}
\end{table*}

We define the reaction embedding as $\mathrm{\textbf{z}_{React}} = \sum  \mathrm{\textbf{z}_{products}} - \sum  \mathrm{\textbf{z}_{reactants}}$. For the formation of \ce{BaTiO3} from its binaries, we observe consistency across models ($\sim>0.7$ c-sim, except the Orb-v3-con model). Embedding compatibility can support zero-shot model stitching.
This allows us to algebraically substitute the product state representation of one model with that of another, treating them as compatible vectors within the shared geometry.
We constructed a hybrid reaction embedding, $\mathrm{\textbf{z}_{React-stitch}}$, by pairing reactant embeddings from MACE-large with product embeddings from other models. As detailed in Table 1, inter-model compatibility is high. MACE-MP-0 variants show strong agreement ($>0.88$). Surprisingly, Orb-v3-con-omat exhibits higher stitchability with MACE-large (0.82) than the other OMat24-trained models. This demonstrates that models trained on non-overlapping datasets (MPtrj vs. OMat24) can be algebraically combined to yield geometrically reasonable embeddings, opening potential routes for modular reuse of pre-trained foundation potentials.

\begin{figure}[H]
\centering
\includegraphics[width=1\linewidth]{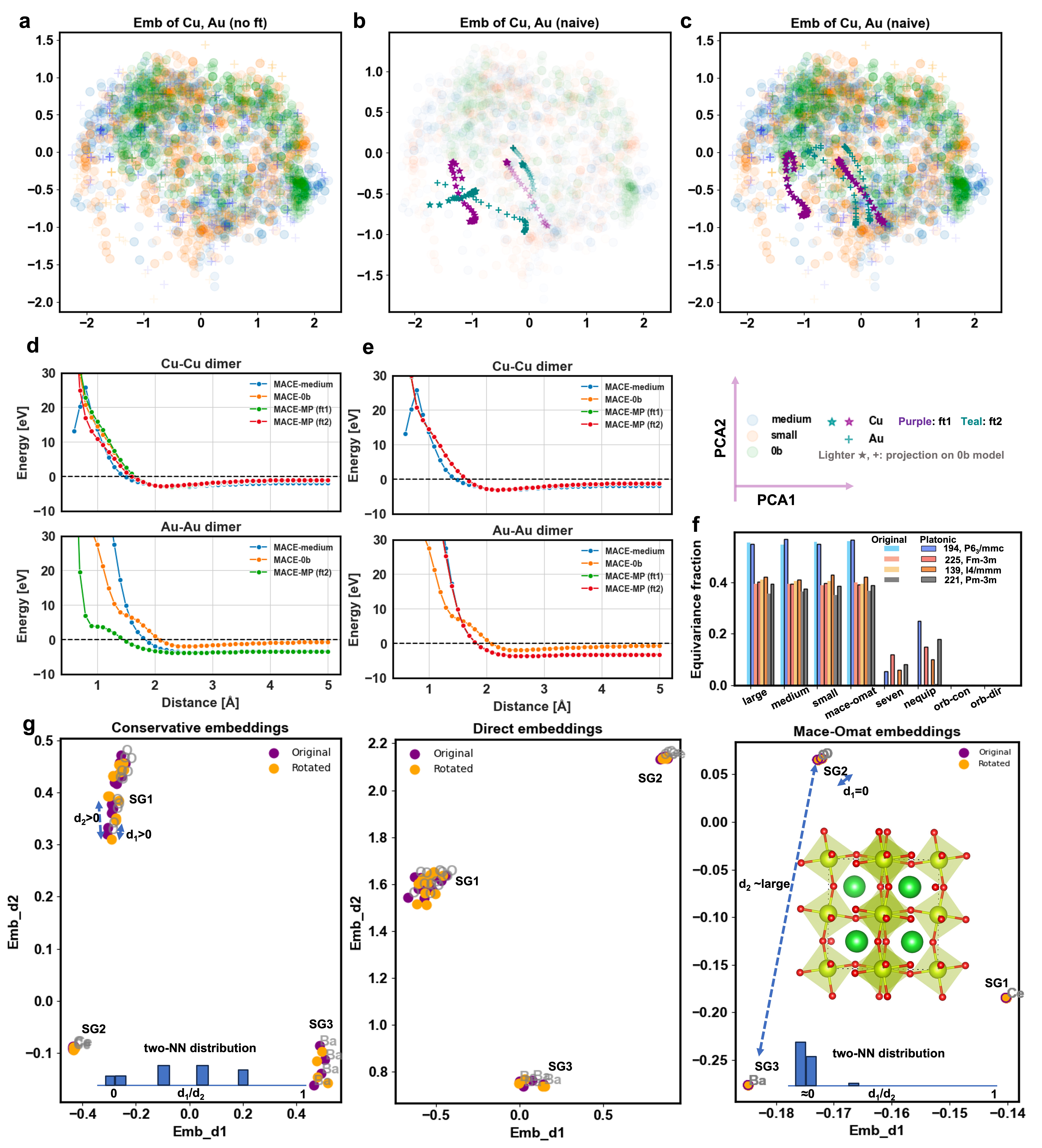}
\caption{\textbf{Diagnostic applications via Platonic embeddings.}
(a) Initial PCA projection of Cu (stars) and Au (crosses) embeddings across MACE models.
(b) Naive fine-tuning (ft1) on additional Cu data causes embeddings for unseen Au atoms to collapse toward the origin (purple trajectories).
(c) Multi-head fine-tuning (ft2) keeps unseen Au embeddings stable near their pre-trained positions (teal trajectories).
(d) Naive embedding collapse results in erroneous, repulsive potentials for Au–Au dimers (green line).
(e) This strategy preserves correct Au–Au physics while selectively adapting Cu–Cu interactions to the target (red line). 
(f) The equivariant fraction of embeddings from different models based on two-nearest-neighbour analysis. 
(g) Rotational sensitivity test on \ce{BaCeO3}: Equivariant models preserve embeddings under rotation; Orb-v3 models do not. }
\label{fig:finetune}
\end{figure}

Finally, this framework serves as a diagnostic tool. In the context of transfer learning (Fig. 5a-e), it visualises the trajectory of atomic embeddings, distinguishing between catastrophic forgetting, where naive fine-tuning causes the weights of unseen species (e.g. Au) to collapse, whereas multi-head strategies align specific interactions (e.g. Cu–Cu) while preserving pre-trained knowledge. 

Beyond training stability, the proposed framework also highlights fundamental limitations in architectures that do not maintain strict equivariance in their learned representations. We adopt the two-nearest-neighbour (two-NN) approach \cite{Facco2017} to quantify effective dimension reduction arising from symmetry equivalence. Considering a reference atom (O) as an example in Fig. 5g, the ratio $d_1$/$d_2$—where $d_1$ and $d_2$ denote the first and second nearest-neighbour distances in representation space—serves as a metric for how the model distinguishes (or fails to distinguish) symmetry-equivalent sites. For perfectly equivariant representations, symmetry-related atoms collapse onto the same point, leading to $d_1$=0. By measuring the fraction of atoms with $d_1$=0, we estimate each model’s capacity to preserve equivariant structure in its learned embeddings.

As shown in Fig. 5f, all equivariant MACE models maintain a consistent level of detected equivalent points across the four representative space groups studied (P$6_3$/mmc, Fm$\bar{3}$m, I4/mmm, Pm$\bar{3}$m), indicating robust preservation of equivariant representations. In contrast, SevenNet although incorporating equivariant e3nn layers, does not detect $d_1$=0 equivariance in their original embeddings here. This may be stem from the architectural approximation inherent in its truncation at $l_{max}$. NequIP, conversely, employs a stringent $\text{e}3\text{nn}$ numerical tolerance of $1e^{-8}$ to $1e^{-10}$ using $\text{float}64$, which improves its numerical precision. This disparity implies that architectural and implementation choices significantly affect numerical performance. However, as detailed in Table S6, we found that both SevenNet and NequIP embeddings could recover their equivariance close to MACE models when assessed at a given numerical tolerance of $1e^{-6}$. 

On the other hand, transformation into Platonic space enables more robust detection of equivariant structure in models with numerical precision limitations, evidenced by the partial re-identification of equivalence clusters  for SevenNet and NequIP models ($\sim15-30\%$, Fig. 5f). This suggests the Platonic representation could potentially benefit from the removal of numerical noise via the projection process. 

For comparison, the non-equivariant Orb models fail to identify any equivalent atoms in either their raw or Platonic embeddings, deviating from physically expected symmetry relations regardless of the numerical accuracy applied, indicating that Platonic transformation cannot restore equivariance that was fundamentally missing in the architecture's learning phase. This failure directly manifests in downstream predictions: as illustrated in Fig. 5g, rotating the $\ce{BaCeO3}$ structure causes the Orb-v3 embeddings to diverge rather than remain invariant, a clear instance of symmetry breaking. This violation propagates into the force field, producing qualitatively incorrect phonon dispersions, including imaginary branches (Fig. S8). 


\section{Conclusions}

We have established that foundation interatomic potentials, despite architectural heterogeneity, disjoint training sets, and distinct inductive biases, exhibit a shared latent geometry. 
By aligning embeddings via an anchor-based projection, we unified these disparate models into a common metric space that preserves chemical periodicity, enables cross-model comparability, and exposes structural invariants previously inaccessible in model-specific coordinates.
We further demonstrated that the Platonic geometry supports meaningful embedding arithmetic for materials, polymorphs, and chemical reactions, effectively allowing zero-shot model stitching. Simultaneously, it serves as a diagnostic tool, revealing geometric distortions in this space that can signal physical prediction failures, such as symmetry breaking in non-equivariant architectures.

Beyond an immediate utility in model comparison, our results suggest a fundamental principle that as physical models in AI for materials scale, their representations may approach compatible geometric organizations that reflect common structure learned from physical constraints. The Platonic framework provides a practical mechanism to probe this convergence and a pathway to interoperable scientific models. We anticipate that future architectures may benefit from considering representational compatibility alongside performance metrics, particularly for enabling model reuse, interoperability and interpretablility.

\section{Methods}

\subsection{Definition of the Platonic representation}
We define the original representation of a trained machine learned potential ($M$) as the direct extraction of embeddings $\mathbf{e}_i = \phi_i(x)$. Let $\mathcal{M}=\{M_1, M_2, \ldots, M_N\}$ be a collection of pretrained MLIPs. Each model $M_i$ defines an embedding function:

\[
\phi_i : \mathcal{X} \rightarrow \mathbb{R}^{d},
\]mapping atomic environments $x \in \mathcal{X}$ to latent vectors $\mathbf{e}_i$.
A canonical embedding, $\mathbf{z}_i(x) \in \mathbb{R}^{K}$ is obtained by projecting each model’s specific embeddings onto a model-independent anchor space using cosine similarity:
\[
\mathbf{z}_i(x) = 
\Big[
\cos(\phi_i(x), \mathbf{a}_1), \;
\cos(\phi_i(x), \mathbf{a}_2), \;
\ldots, \;
\cos(\phi_i(x), \mathbf{a}_K)
\Big]^\top,
\]
where $\{\mathbf{a}_k\}_{k=1}^{K}$ is an anchor set sampled from the union of model embeddings and embedded into a unified metric space. This representation satisfies four key properties:

\begin{enumerate}
\item \textbf{Coordinate invariance.} For any orthogonal linear transformation $Q_i$ applied to the original space, $T(Q_i \phi_i(x)) \approx T(\phi_i(x))$, implying the transformation eliminates architecture-dependent coordinate rotations.

\item \textbf{Model universality.} For embeddings of the same physical environment across different models $i$ and $j$:
\[
\| \mathbf{z}_i(x) - \mathbf{z}_j(x) \| \ll \| \phi_i(x) - \phi_j(x) \|,
\]demonstrating convergence toward a shared latent geometry.

\item \textbf{Physical latent consistency.} Distances in the Platonic space reflect intrinsic physical similarity:
\[
\| \mathbf{z}(x) - \mathbf{z}(x') \| \propto \text{similarity of atomic environments}.
\]

\item \textbf{Algebraic compatibility.} Structure- or reaction-level embeddings satisfy $\mathbf{z}_{M_i}(X) \approx \mathbf{z}_{M_j}(X)$, enabling cross-model optimal transport and arithmetic.
\end{enumerate}

\subsection{Dataset and embedding extraction}

We utilised 27,136 structures from the MP-20 training dataset \cite{Xie2021CrystalDV} as the target set for embedding extraction. While MACE models provide a native \textit{get\_descriptor} function, other architectures required custom interfaces to access the latent layers. An extraction script was used to generate a total of 282,847 atomic embeddings per model. All extracted model-wise embeddings have been archived to facilitate anchor set generation and reproducibility in Zenodo DOI: 10.5281/zenodo.17721681.

\subsection{DIRECT sampling strategy}

To construct the anchor set, we applied the DIRECT sampling strategy\cite{Qi2024}. This method is designed to maximise coverage of the chemical latent space. To maintain consistency, we used the distribution of embeddings from the MACE-small model as the reference manifold. Anchors were selected by applying Birch clustering to the reference distribution (target $K=100$ clusters), followed by stratified selection to identify centroids that maximise the Silhouette score. The indices of these selected atoms were then used to retrieve the corresponding embeddings from all other models, ensuring that the anchors represent physically identical atomic environments across architectures.

\subsection{Dummy model construction}

To control for architectural inductive biases, we constructed a `Dummy' baseline. This model shares the architecture of MACE-small but is initialised with random weights ($w$) and biases ($b$) drawn from a standard normal distribution. This results in a model with 3,847,696 randomised parameters that has seen no training data, serving as a negative control to verify that the Platonic geometry arises from learned physical correlations rather than architectural priors.

\subsection{Similarity and distance metrics}
To quantify representational alignment, we computed the following metrics:

\textbf{Procrustes Score.} Measures the Euclidean distance between two representations, $A$ and $B$, after optimal orthogonal alignment. It is computed by minimising the Frobenius norm $\| A - BQ \|_F$ over all orthogonal matrices $Q$. We report the similarity score derived from the residual sum of squares:

\[
\mathrm{Score_{Procrustes}} = 1 - \frac{\| A - BQ^* \|_F^2}{\| A - \bar{A} \|_F^2 + \| B - \bar{B} \|_F^2},
\]where $Q^*$ is the optimal rotation. A high score indicates that the two spaces share the same global topology up to rotation and scaling.

\textbf{Centered Kernel Alignment (CKA).} CKA measures the similarity between the centered Gram matrices of two representations. It captures how similarly the two spaces organise all pairwise relationships between samples, independent of the explicit rotation required by Procrustes analysis.

\textbf{Optimal Transport (OT) Cost.} We compute the normalised OT cost to measure the minimal effort required to transform the distribution of model $A$ into model $B$. We utilise the Sinkhorn algorithm to approximate the Wasserstein distance between the two embedding distributions in the unified Platonic space.

\textbf{Mutual k-Nearest Neighbours (mKNN).} To assess local topological fidelity, we calculate the overlap of local neighbourhoods. For each sample, we identify the set of $k$ nearest neighbours in Model $A$ and Model $B$. The mKNN score is defined as the Jaccard index of these two sets, averaged over all samples.

\section*{Acknowledgements}

We thank Lars Schaaf and Kinga Mastej for useful discussions and suggestions related to embedding analysis and the chemical consequences. 
We are grateful to the UK Materials and Molecular Modelling Hub for computational resources, which is partially funded by EPSRC (EP/T022213/1, EP/W032260/1 and EP/P020194/1).
We thank the EPSRC for support via the AI for Chemistry: AIchemy hub (EPSRC grant EP/Y028775/1 and EP/Y028759/1).

\section*{Author contributions statement}
The author contributions have been defined following the CRediT system.
Z.L.: Conceptualization, Investigation, Formal analysis, Methodology, Visualization, Writing – original draft. 
A.W.: Conceptualization, Methodology, Writing – review \& editing.

\section*{Data access statement}
All of the foundation models analysed are openly available. The codes used to perform the transformations and analysis are available in an open-source repository on \url{https://github.com/WMD-group/PlatonicRep}.

\bibliographystyle{unsrt} 
\bibliography{sample.bib}

\end{document}



%
%

\section{Non-linear projection of embeddings}
We have down-selected 5000 embeddings randomly from the full phenomenological embedding space and adopted the UMAP visulization. No clear Platonic nature being captured by the non-linear projection of embeddings. 
\begin{figure}[H]
\includegraphics[width=\textwidth]{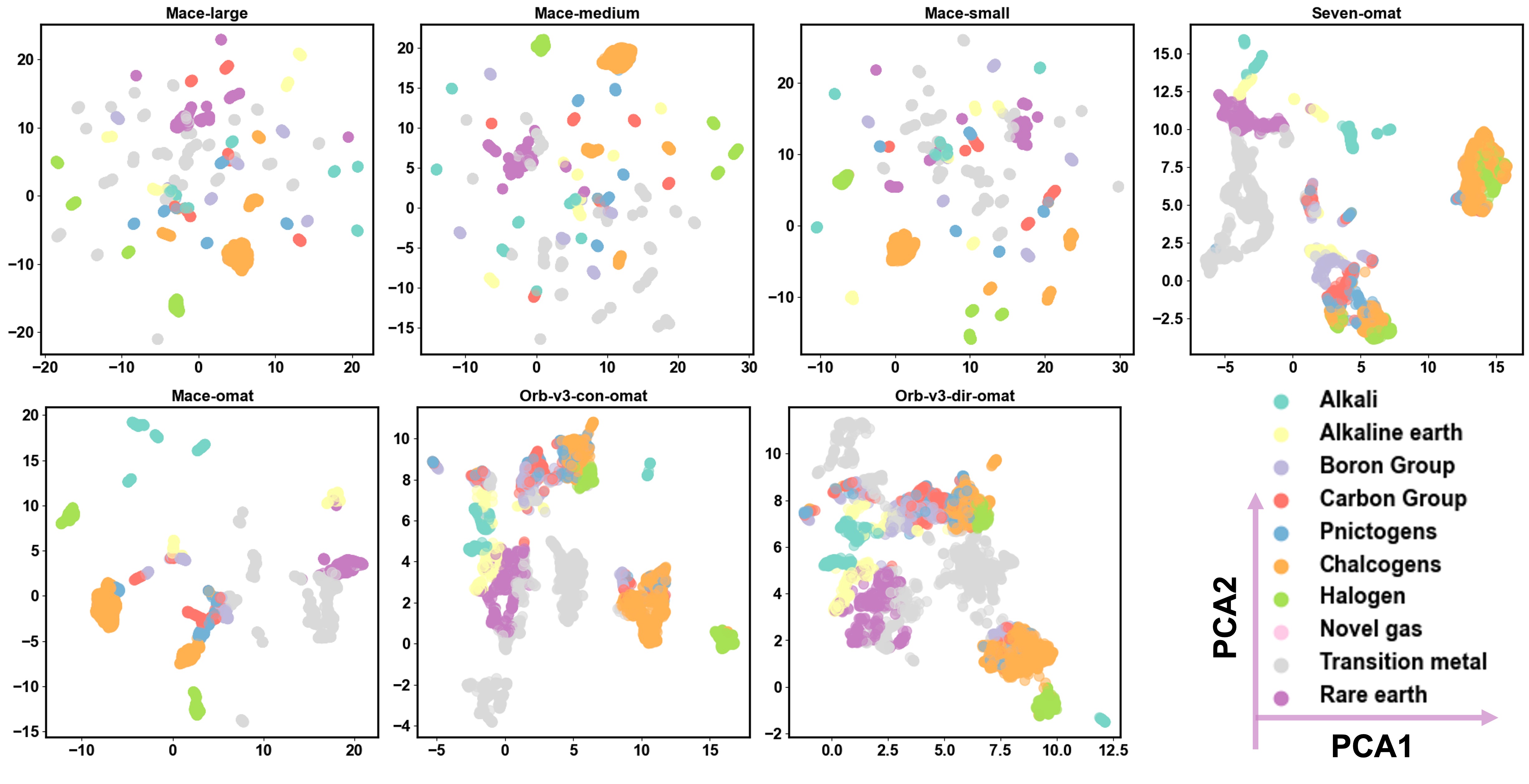}
\caption{UMAP visulization of the seven chosen MLIPs, with random\_state=42, n\_neighbors=30.}
\label{s1}
\end{figure}

\section{Justification of Anchor selection strategy}
We compared the diversity and clustering of anchor sets obtained both from random sampling and the DIRCET Sampling strategy.

\begin{table*}
\caption{Silhouette scores and pairwise distances (in parentheses) for anchor sets selected using different strategies. For random sampling, the values correspond to runs with random seeds 0, 42, and 12345. For DIRECT Sampling, the values correspond to threshold\_init settings of 0.1, 0.2, and 0.3. When evaluating clusterability using KMeans, the number of target clusters is set to 10, matching the number of labeled elemental groups. A lower Silhouette score indicates a less clusterable anchor set, whereas a larger pairwise distance indicates a more diverse anchor set.}
\label{table1}
\setlength{\arrayrulewidth}{0.4mm}
\begin{tabular}{cccc}
\hline
Size of Anchor set & Random Sampling & DIRECT Sampling 
\\ 
\hline
100   & 0.09 (0.4), 0.14(0.1), 0.14(0.45) & 0.14(0.62), 0.14(0.47), \textbf{0.06(1.73)} 
\\
200  & 0.17(0.4), 0.17(0.1), 0.16(0.43)  & \textbf{0.08(1.47)}, 0.06(0.84), 0.09(0.79)
\\
400  & 0.12(0.1), 0.14 (0.1), 0.13(0)  & 0.07(0.59), \textbf{0.09(0.74)}, 0.10(0.31)
\\
\hline
\end{tabular}
\end{table*}

\section{Robustness of anchor-based transformation}

The platonic transformation of seven chosen MLIPs and the dummy model corresponding to different sizes of anchor sets. 
\begin{figure}[H]
\includegraphics[width=\textwidth]{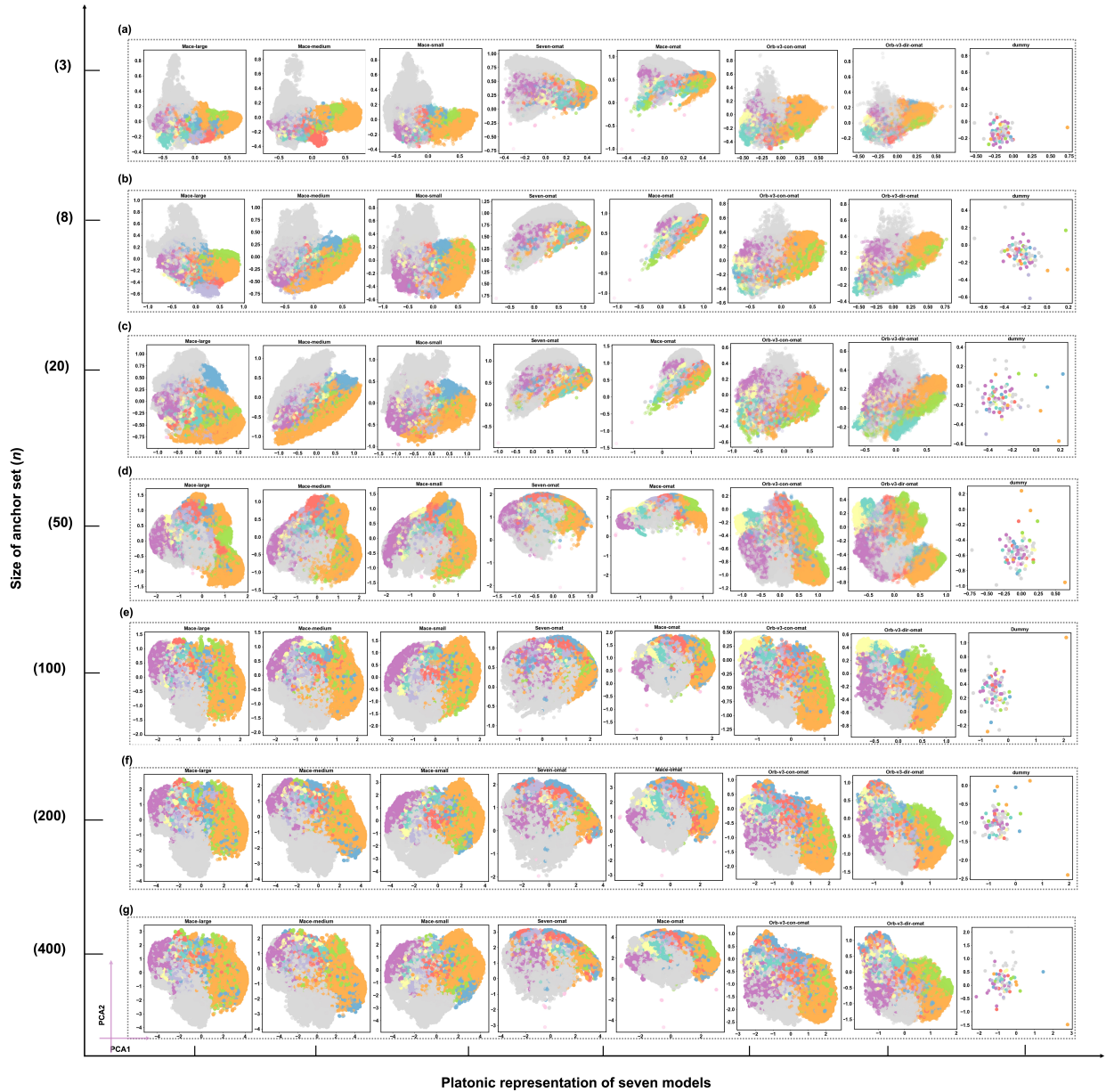}
\caption{Seven chosen MLIPs and the dummy model in the unifed latent space defined by different sizes of anchor sets, selected with the DIRECT sampling strategy.}
\label{s1}
\end{figure}

\begin{figure}[H]
\includegraphics[width=\textwidth]{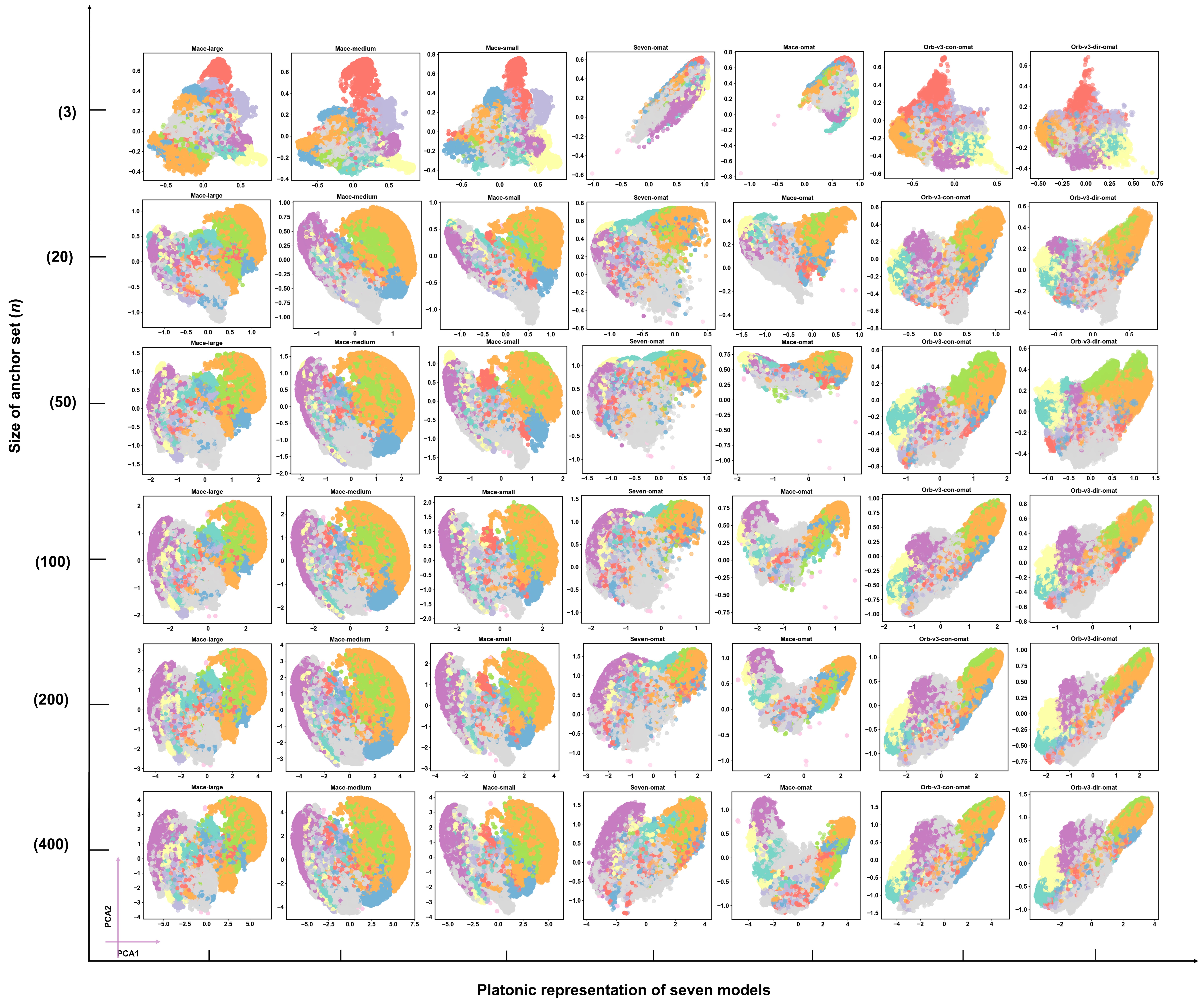}
\caption{Seven chosen MLIPs in the unifed latent space defined by different sizes of anchor sets, selected with the random sampling.}
\label{s1}
\end{figure}

\begin{figure}[H]
\includegraphics[width=\textwidth]{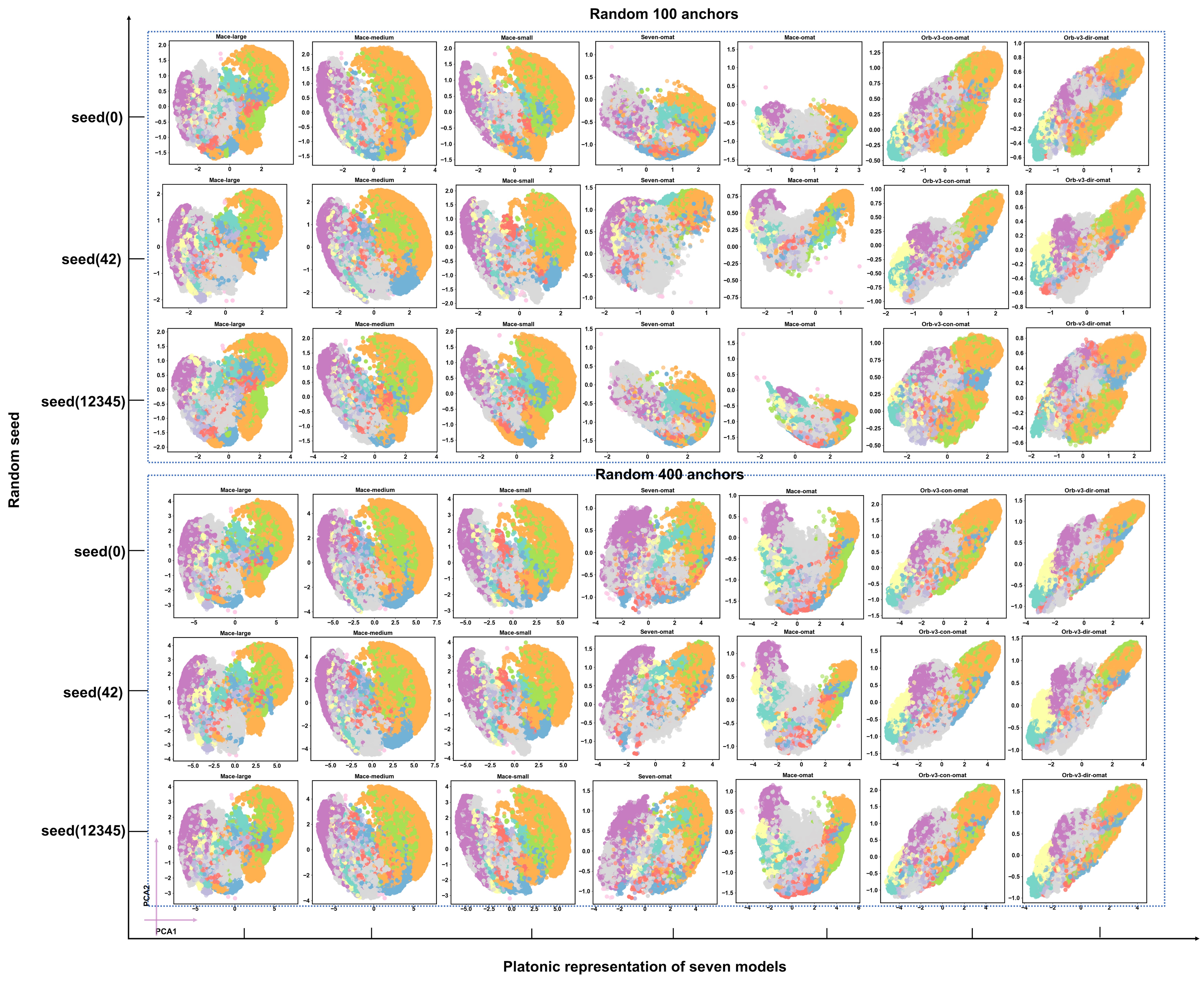}
\caption{The robustness of anchor-based transformation tested againt different random seeds.}
\label{s1}
\end{figure}

\section{Evaluate the converged form of unified representations}
To quantify the clustering score of converged representations after transformation for both anchor selection strategies, the Silhouette scores ($S$) and mean pairwise distances ($D$) were computed, where smaller mean pairwise distances ($D$) means more compact clustering after transformation. 

Results in Table S2 show that, DIRECT sampling generated more compact clustering with smaller $D$ in contrast to the values from random sampling. Both converged representations show similar Silhouette scores ($S$) for each model.

\begin{table*}
\caption{Silhouette scores ($S$) and mean pairwise distances ($D$) for the converged transformed embeddings using random and DIRECT sampling. Using random seed 42 to down selected 5000 embeddings.}
\label{table1}
\setlength{\arrayrulewidth}{0.4mm}
\begin{tabular}{cccccccc}
\hline
&large & medium & small & 7omat & Mace-omat & Orb-v3-con-omat & Orb-v3-dir-omat 
\\ 
\hline
$S$(Random)&0.30 &0.28& 0.29&0.22 & 0.28& 0.16&0.12
\\
$S$(DIRECT)&0.28&0.26&0.27&0.23&0.26& 0.15&0.11
\\
$D$(Random)&3.06 &3.11& 2.71&2.18 & 2.20& 2.28&1.80
\\
$D$(DIRECT)&\textbf{2.60} &\textbf{2.67}& \textbf{2.34}&\textbf{2.10} & \textbf{2.08}& \textbf{2.00} & \textbf{1.58}
\\
\hline
\end{tabular}
\end{table*}

\section{Platonic representation holds for extended models} 
\begin{figure}[H]
\includegraphics[width=\textwidth]{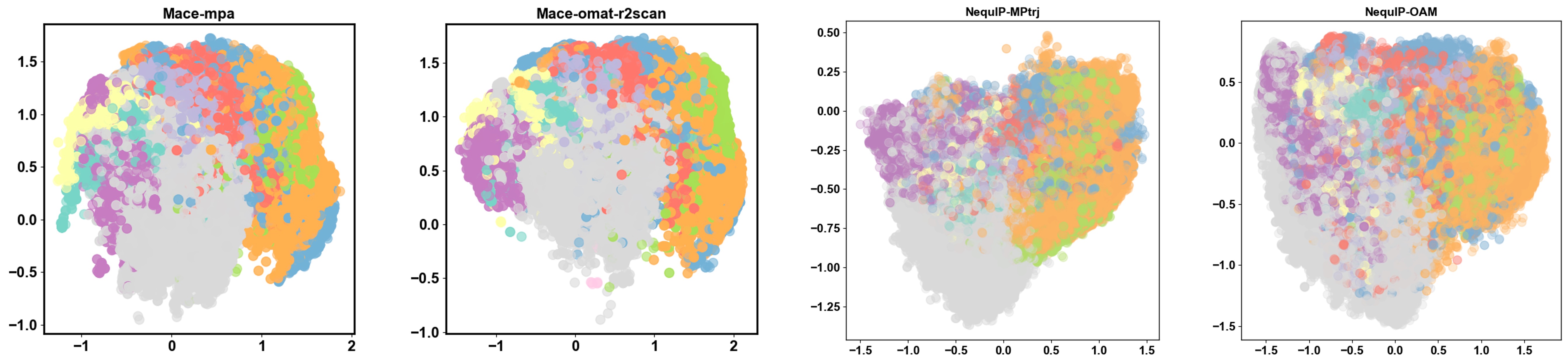}
\caption{The unified representation for foundational MACE-MPA-0, MACE-omat-r2scan, NequIP-MPtrj and NequIP-OAM models.}
\label{s1}
\end{figure}

\section{Supplement metrics}
We have used the mKNN to measure the local arrangement of embeddings to measure the neighborhood overlap for each sample; for the CKA measurement, it is global and measures the overall pairwise relationship alignment, therefore, this measure requires dimensionality match; and it is used to compare with the Procrustes score, which is also global and measure the best rotation/alignment distance.

\begin{figure}[H]
\includegraphics[width=\textwidth]{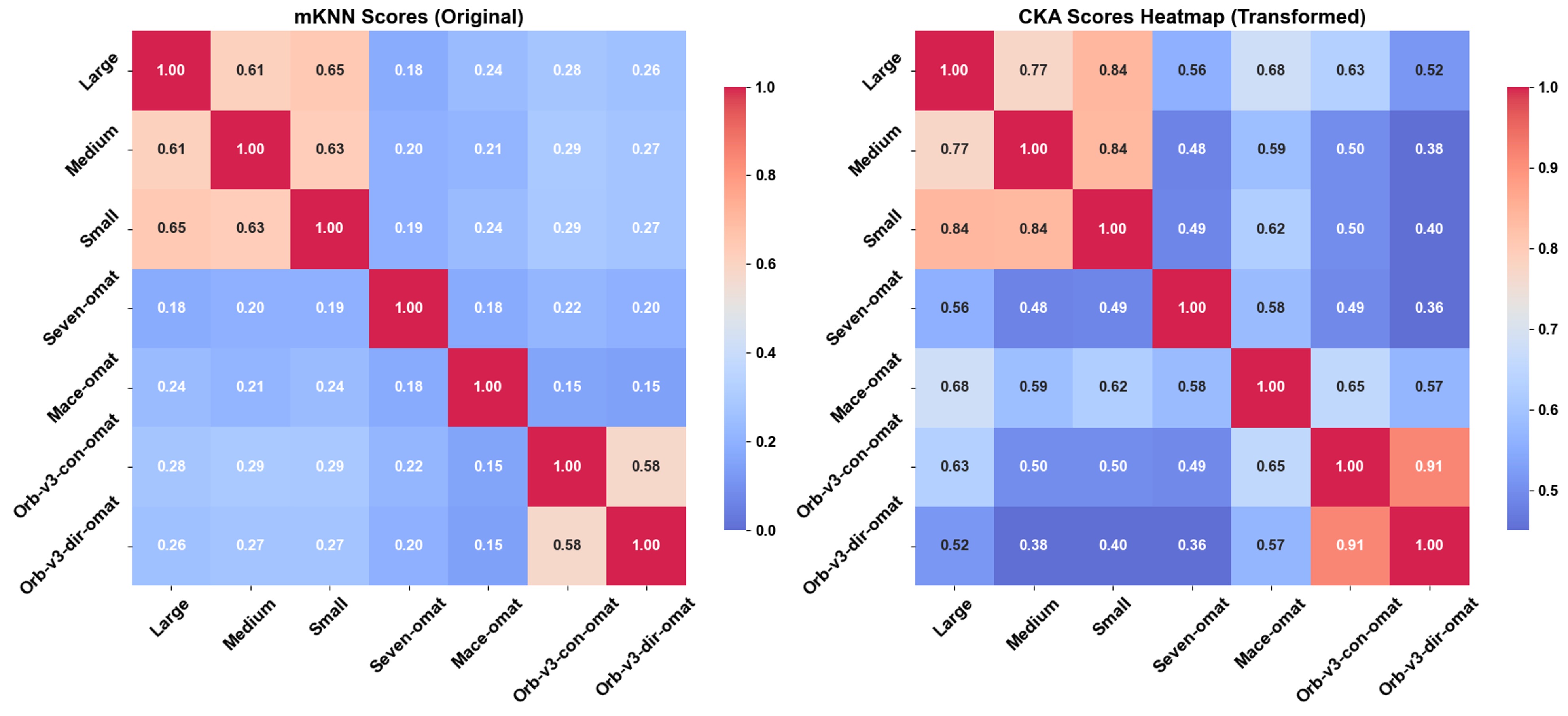}
\caption{(Left) mKNN performed on the original embedding space. (Right) The unified representation enables the CKA measurement, which shows consistent results as the $\mathrm{Score_{Procrustes}}$.}
\label{s1}
\end{figure}

\begin{figure}[H]
\includegraphics[width=\textwidth]{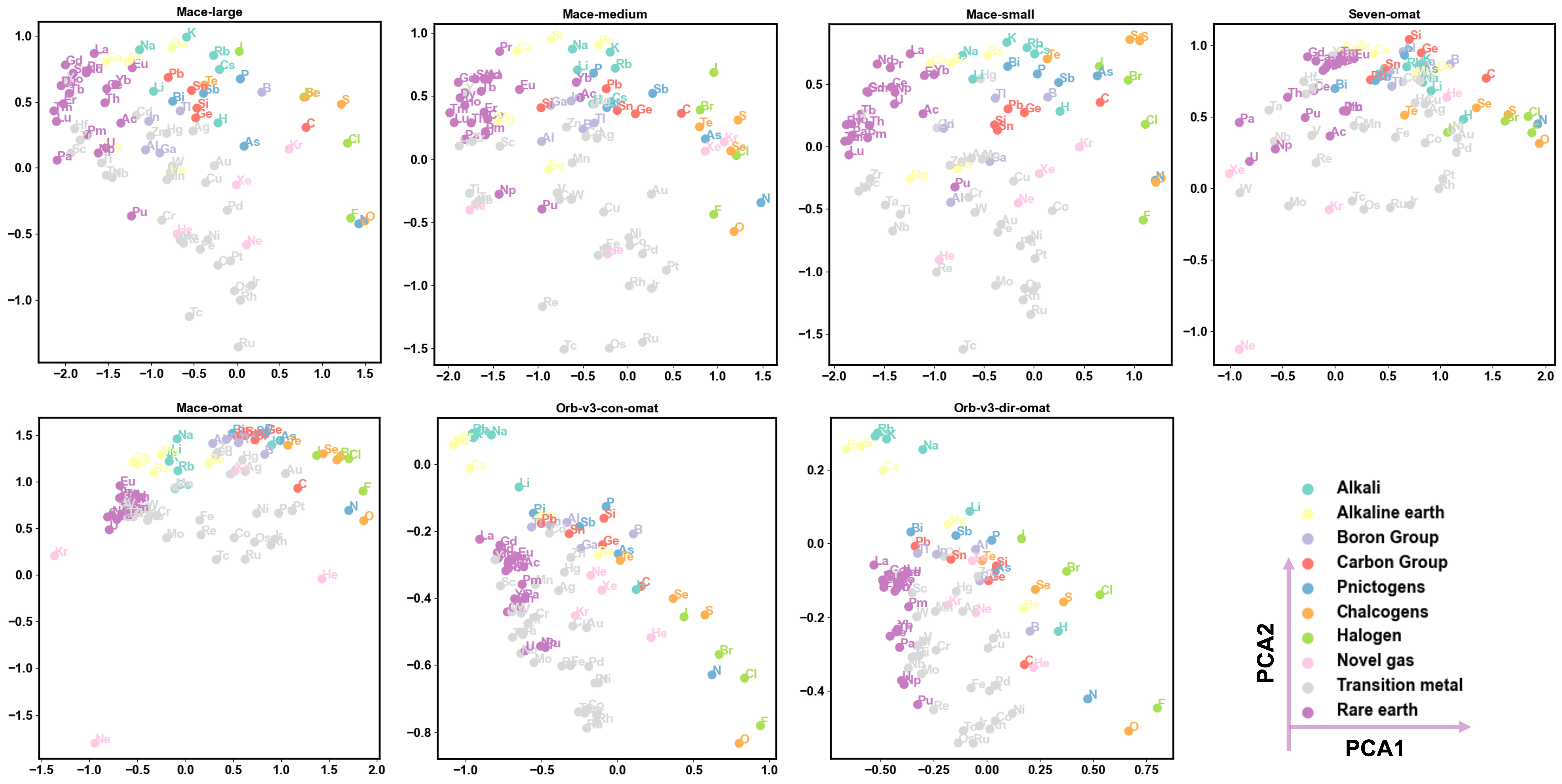}
\caption{Element-level embeddings projected into the unified space ($K$=100 anchors) reveal consistent periodic clustering across all seven models.}
\label{s1}
\end{figure}

\begin{figure}[H]
\includegraphics[width=\textwidth]{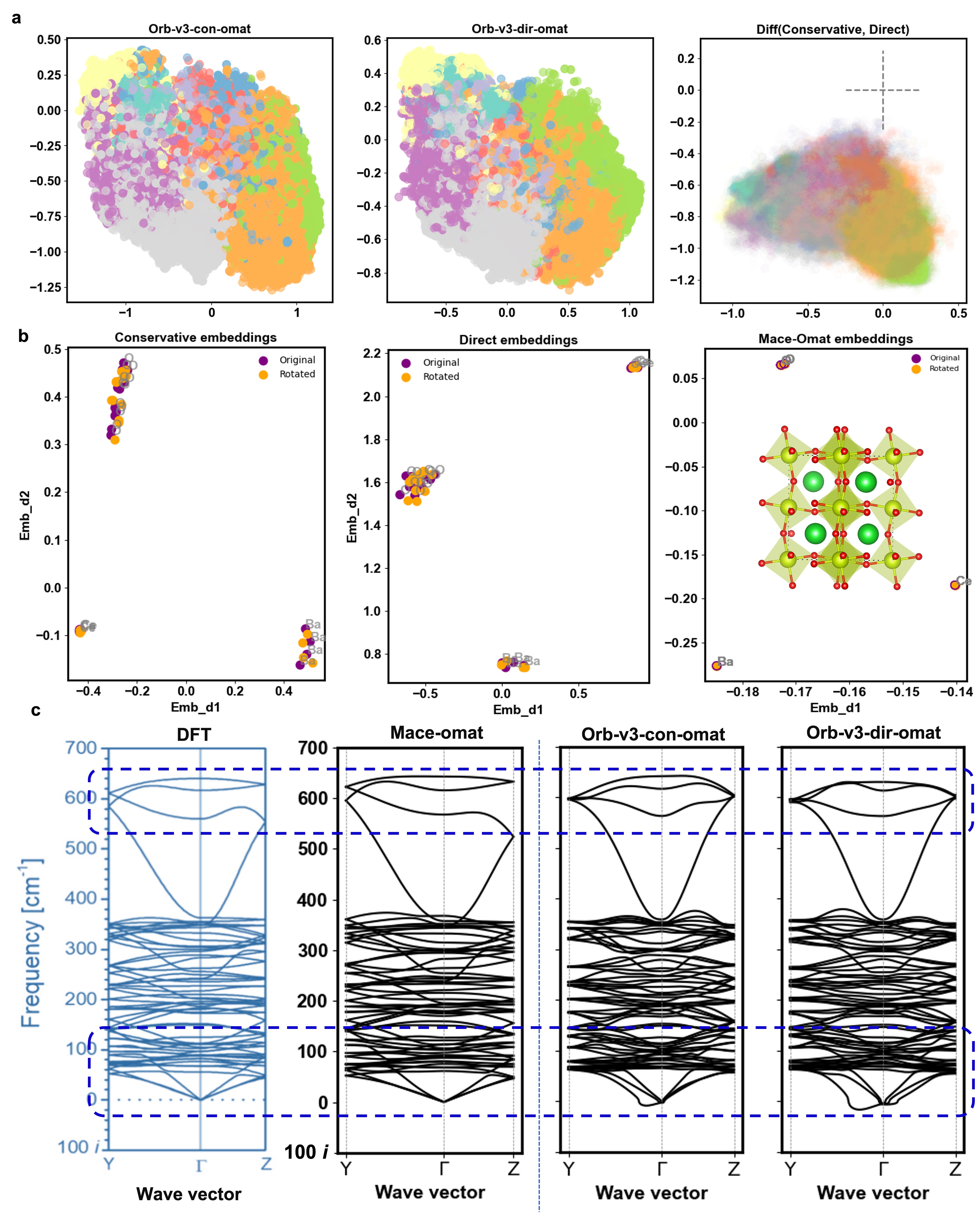}
\caption{\textbf{Detecting symmetry breaking.} (a) PCA projection of the difference vector between Orb-v3 conservative and non-conservative embeddings. (b) Rotational sensitivity test on \ce{BaCeO3}: Equivariant models preserve embeddings under rotation; Orb-v3 models do not. (c) These representational errors propagate to physical predictions, resulting in incorrect phonon dispersions and imaginary acoustic modes close to $\Gamma$.}
\label{s1}
\end{figure}

\section{Unified embedding arthimetic}
$\mathrm{\mathbf{Emb_{Mater}}}$: Extended set of materials and the $l$s of their $\mathrm{Emb_{Mater}}$, where scaling factor1 is the $l$ from the Seven-omat relative to that from the Mace-large model and the scaling factor2 is the $l$ from the Mace-omat relative to that from the Mace-large model.
\begin{table*}
\caption{Scaling factor between Mace-omat, Seven-omat models relative to the Mace-large models.}
\label{table1}
\setlength{\arrayrulewidth}{0.4mm}
\begin{tabular}{cccccccc}
\hline
&Seven-omat & Mace-omat & large & scaling factor1 & scaling factor2  
\\ 
\hline
\ce{Na3MnCoNiO6}&4.50 &5.44& 1.14&2.94 & 3.56
\\
\ce{Nd(Al2Cu)4}&6.17 &6.32& 1.66&2.88 & 2.96
\\
\ce{LiMnIr2}&5.17 &6.35& 1.50&3.00 & 3.68
\\
\ce{LiCSN}&5.45 &6.36& 1.49&3.24 & 3.78
\\
\ce{La2EuS4}&5.71 &6.31& 1.89&2.71 & 2.99
\\
\ce{Yb3Ga9Pt2}&6.08 &6.31&1.93&2.79 & 2.90
\\
\ce{NaLiCO3}&5.34 &5.95& 1.56&4.06 & 4.52
\\
\ce{Na2BiO3}&5.09 &5.68& 1.59&3.23 & 3.61
\\
\hline
\end{tabular}
\end{table*}

$\mathrm{\mathbf{Emb_{React}}}$: Extended set of hypothetical reactions to check the cross-model consistency of $\mathrm{Emb_{React}}$ and the zero-shot stitching compatibility between all the other models with the Mace-large model. 

\begin{table*}[h!]
\caption{R1: BaO + \ce{ZrO2} $\rightarrow$ \ce{BaZrO3}; R2: 2CsF + \ce{InF3} + AgF $\rightarrow$ \ce{CsInAgF6} + Cs; R3: CsCl + \ce{SnCl2} $\rightarrow$ \ce{CsSnCl3}, where for each reaction we list the reaction embeddings from intra-model and stitched model with Mace-small, with respect to ($l$, c-sim). }

\label{table1}
\setlength{\arrayrulewidth}{0.4mm}
\begin{tabular}{ccccccc}
\hline
MLIPs & R1 & Stitch-R1 & R2 & Stitch-R2 & R3 & Stitch-R3
\\ 
\hline
Mace-large   & 1.31, 1 & 1.78, 1 & 3.57, 1 & 3.34, 1 & 1.86, 1 & 1.64, 1
\\
Mace-medium  & 1.11, 0.82 & 2.01, 0.91 & 3.08, 0.85 & 4.51, 0.95 & 1.48, 0.86 & 2.15, 0.95
\\
Mace-small  & 1.30, 0.87 &1.30, 0.84 & 3.16, 0.92 &3.16, 0.93 & 1.74, 0.97 &1.74, 0.96
\\
Seven-omat & 5.43, 0.76 &3.44, 0.34 &16.66, 0.73  & 3.85, 0.51 & 5.72, 0.82 &3.06, 0.17
\\ 
Mace-omat & 5.71, 0.75 &4.16, 0.39 &14.73, 0.72  &4.60, 0.35 & 5.45, 0.86 &3.24, 0.32
\\ 
Orb-v3-con-omat & 1.17, 0.51 &2.72, 0.84 & 2.80, 0.58 &5.97, 0.91 & 0.91, 0.33 &3.37, 0.97
\\ 
Orb-v3-dir-omat & 1.48, 0.67 & 2.27, 0.63 & 4.14, 0.74 & 4.81, 0.91& 1.60, 0.86 &3.56, 0.97
\\
\hline
\end{tabular}
\end{table*}

\section{Re-discovered equivariant atoms by Platonic representation}

We have collected all the embeddings pairs that have distances below 1E-8 as the ground truth of equivariant pairs. 

\renewcommand{\arraystretch}{1.5}
\begin{table*}[h!]
\caption{Count of equivariant atoms based on brute force distance calculation, denoted as $\mathrm{\frac{Count_{Original}.}{Count_{Platonic}}}$. Accuracy: 1E-8}
\label{table1}
\setlength{\arrayrulewidth}{0.4mm}
\begin{tabular}{ccccccccc}
\hline
space group & large & medium & samll & Mace-omat & Seven & NeuqIP & orb-con & orb-dir
\\ 
\hline
221& $\mathrm{\frac{2674}{2674}}$ & $\mathrm{\frac{2674}{2674}}$ & $\mathrm{\frac{2674}{2674}}$ & $\mathrm{\frac{2674}{2674}}$ & $\mathrm{\frac{3}{3}}$ & $\mathrm{\frac{2132}{2}}$ & $\mathrm{\frac{0}{0}}$ & $\mathrm{\frac{0}{0}}$
\\
225& $\mathrm{\frac{22920}{22922}}$ & $\mathrm{\frac{22945}{22925}}$ & $\mathrm{\frac{22920}{22928}}$ & $\mathrm{\frac{22942}{22946}}$ & $\mathrm{\frac{2}{2}}$ & $\mathrm{\frac{15720}{10}}$ & $\mathrm{\frac{0}{0}}$ & $\mathrm{\frac{0}{0}}$
\\
139& $\mathrm{\frac{4878}{4896}}$ & $\mathrm{\frac{4896}{4896}}$ & $\mathrm{\frac{4877}{4896}}$ & $\mathrm{\frac{4881}{4881}}$ & $\mathrm{\frac{4}{4}}$ & $\mathrm{\frac{3625}{0}}$ & $\mathrm{\frac{0}{0}}$ & $\mathrm{\frac{0}{0}}$
\\
194& $\mathrm{\frac{9248}{9252}}$ & $\mathrm{\frac{9248}{9252}}$ & $\mathrm{\frac{9247}{9252}}$ & $\mathrm{\frac{9248}{9252}}$ & $\mathrm{\frac{189}{189}}$ & $\mathrm{\frac{9410}{2}}$ & $\mathrm{\frac{0}{0}}$ & $\mathrm{\frac{0}{0}}$
\\
\hline
\end{tabular}
\end{table*}

\renewcommand{\arraystretch}{1.5}
\begin{table*}[h!]
\caption{Count of equivariant atoms based on brute force distance calculation, denoted as $\mathrm{\frac{Count_{Original}.}{Count_{Platonic}}}$. Accuracy: 1E-6}
\label{table1}
\setlength{\arrayrulewidth}{0.4mm}
\begin{tabular}{ccccccccc}
\hline
space group & large & medium & samll & Mace-omat & Seven & NeuqIP & orb-con & orb-dir
\\ 
\hline
221& $\mathrm{\frac{2677}{2678}}$ & $\mathrm{\frac{2678}{2677}}$ & $\mathrm{\frac{2677}{2678}}$ & $\mathrm{\frac{2678}{2678}}$ & $\mathrm{\frac{2663}{2673}}$ & $\mathrm{\frac{2733}{2629}}$ & $\mathrm{\frac{0}{0}}$ & $\mathrm{\frac{0}{0}}$
\\
225& $\mathrm{\frac{23308}{23325}}$ & $\mathrm{\frac{23327}{23318}}$ & $\mathrm{\frac{23288}{23308}}$ & $\mathrm{\frac{23334}{23350}}$ & $\mathrm{\frac{22975}{23285}}$ & $\mathrm{\frac{27162}{23379}}$ & $\mathrm{\frac{0}{0}}$ & $\mathrm{\frac{0}{0}}$
\\
139& $\mathrm{\frac{4993}{5011}}$ & $\mathrm{\frac{5027}{5013}}$ & $\mathrm{\frac{4990}{5004}}$ & $\mathrm{\frac{5025}{5076}}$ & 
$\mathrm{\frac{4867}{4988}}$ & $\mathrm{\frac{5614}{5193}}$ & $\mathrm{\frac{0}{0}}$ & $\mathrm{\frac{0}{0}}$
\\
194& $\mathrm{\frac{9831}{10670}}$ & $\mathrm{\frac{10749}{10770}}$ & $\mathrm{\frac{9668}{10864}}$ & $\mathrm{\frac{9820}{11151}}$ & $\mathrm{\frac{9134}{10191}}$ & $\mathrm{\frac{15304}{13617}}$ & $\mathrm{\frac{0}{0}}$ & $\mathrm{\frac{0}{0}}$
\\
\hline
\end{tabular}
\end{table*}

\renewcommand{\arraystretch}{1.5}
\begin{table*}[h!]
\caption{Count of re-discovered equivariant atoms by Platonic representation, denoted as $\mathrm{\frac{Count_{Original}}{Count_{Platonic}}}$ with the two-NN algorithm.}
\label{table1}
\setlength{\arrayrulewidth}{0.4mm}
\begin{tabular}{ccccccccc}
\hline
space group & large & medium & samll & Mace-omat & Seven & NeuqIP & orb-con & orb-dir
\\ 
\hline
221& $\mathrm{\frac{1247}{1389}}$ & $\mathrm{\frac{1283}{1316}}$ & $\mathrm{\frac{1228}{1354}}$ & $\mathrm{\frac{1289}{1364}}$ & $\mathrm{\frac{0}{286}}$ & $\mathrm{\frac{14}{628}}$ & $\mathrm{\frac{0}{0}}$ & $\mathrm{\frac{0}{0}}$
\\
225  & $\mathrm{\frac{9098}{9279}}$ & $\mathrm{\frac{9065}{9091}}$ & $\mathrm{\frac{8941}{9160}}$ & $\mathrm{\frac{9184}{9031}}$ & $\mathrm{\frac{4}{2470}}$ & $\mathrm{\frac{59}{3433}}$ & $\mathrm{\frac{0}{0}}$ & $\mathrm{\frac{0}{0}}$
\\
139  & $\mathrm{\frac{3356}{3429}}$ & $\mathrm{\frac{3293}{3343}}$ & $\mathrm{\frac{3301}{3500}}$ & $\mathrm{\frac{3202}{3428}}$ & $\mathrm{\frac{0}{492}}$ & $\mathrm{\frac{4}{810}}$ & $\mathrm{\frac{0}{0}}$ & $\mathrm{\frac{0}{0}}$
\\
194  & $\mathrm{\frac{5276}{5212}}$ & $\mathrm{\frac{5190}{5393}}$ & $\mathrm{\frac{5288}{5367}}$ & $\mathrm{\frac{20}{499}}$ & $\mathrm{\frac{13}{2363}}$ & $\mathrm{\frac{14}{628}}$ & $\mathrm{\frac{0}{0}}$ & $\mathrm{\frac{0}{0}}$
\\
\hline
\end{tabular}
\end{table*}

\section{Dummy MACE model}
\begin{lstlisting}[language=Python, caption={Randomizing MACE-small foundation model}]
import torch
from mace.calculators import mace_mp
# Get the foundation model with return_raw_model=True
model = mace_mp(model="small", device="cuda", return_raw_model=True)
# Randomize all weights
for param in model.parameters():
    if param.requires_grad:
        torch.nn.init.normal_(param, mean=0.0, std=0.02)
# Save the randomized model
torch.save(model, "dummy_mace_mp_small_randomized.model")
print(f"Randomized model with {sum(p.numel() for p in model.parameters())} parameters") 
\end{lstlisting}


\clearpage
